\documentclass[twocolumn,aps,floatfix,superscriptaddress,nofootinbib,prb]{revtex4-1}

\usepackage{graphicx}
\usepackage{amsmath,amsthm,amssymb}
\usepackage{bm}
\usepackage{esint} 
\usepackage{titlesec} 
\newlength{\titleskip}
\titleformat{name=\section,numberless}[block]{\large\sffamily\bfseries}{\thesection}{0em}{}
\titlespacing*{name=\section,numberless}{0em}{9pt}{-\parskip}

\usepackage[english]{babel}
\usepackage{ulem}
\usepackage{color}
\definecolor{darkblue}{rgb}{0,0,1}
\definecolor{darkgreen}{rgb}{0,0.60,.2}

\def\DB{\mathbf{D}_\mathrm{B}}
\def\DD{\Delta\mathbf{D}}
\def\rmd{\mathrm{d}}
\def\vq{\mathbf{q}}

\begin{document}

\title{Anderson localization of ultracold atoms: Where is the mobility edge?}

\author{Michael Pasek}
\affiliation{Laboratoire Kastler Brossel, UPMC-Sorbonne Universit\'es, CNRS, ENS-PSL Research University, Coll\`{e}ge de France, 4 Place Jussieu, 75005 Paris, France}
\affiliation{Laboratoire Charles Fabry UMR 8501, Institut d'Optique, CNRS, Univ Paris Sud 11, 2 Avenue Augustin Fresnel, 91127 Palaiseau cedex, France}
\affiliation{Laboratoire Mat\'eriaux et Ph\'enom\`enes Quantiques, Universit\'e Paris Diderot-Paris 7 and CNRS, UMR 7162, 75205 Paris Cedex 13, France}

\author{Giuliano Orso}
\affiliation{Laboratoire Mat\'eriaux et Ph\'enom\`enes Quantiques, Universit\'e Paris Diderot-Paris 7 and CNRS, UMR 7162, 75205 Paris Cedex 13, France}

\author{Dominique Delande}
\affiliation{Laboratoire Kastler Brossel, UPMC-Sorbonne Universit\'es, CNRS, ENS-PSL Research University, Coll\`{e}ge de France, 4 Place Jussieu, 75005 Paris, France}

\begin{abstract}
Recent experiments on non-interacting ultra-cold atoms in correlated disorder have yielded conflicting results regarding the 
so-called mobility edge, i.e. the energy threshold separating  Anderson localized from diffusive states. At the same time, there are theoretical indications
that the experimental data overestimate the position of this critical energy, sometimes by a large amount.
The non-trivial effect of anisotropy in the spatial correlations of experimental speckle potentials  have been put forward 
as a possible cause for such discrepancy. 
Using extensive numerical simulations we show that the effect of anisotropy 
alone is not sufficient to explain the experimental data. In particular, we find
that, for not-too-strong anisotropy, realistic disorder configurations are essentially identical to the isotropic case, modulo a simple
rescaling of the energies.
\end{abstract}


\maketitle

When a wave travels in a random medium, the interference between multiple scattering paths caused by the disorder barriers
can completely stop its diffusion. This phenomenon, known as Anderson localization~\cite{Anderson}, is completely general and 
applies to any kind of waves including light waves in diffusive
media~\cite{Wiersma:Nature1997,Storzer:PRL2006} or in photonic crystals~\cite{Schwartz:Nature2006,Lahini:PRL2008}, 
ultrasound~\cite{Hu:NatPhy2008}, microwaves~\cite{Chabanov:Nature2008} and atomic matter waves~\cite{Billy:Nature2008,Roati:Nature2008}, the latter describing the behavior of 
non-interacting atoms in the low temperature quantum regime.
Localization experiments using cold atomic gases have several advantages over their solid-state counterparts\cite{Katsumoto:FineTuningOfMetalInsulator:JPSJ87,Waffenschmidt:MIT:PRL99}. 
The inhibition of transport may be directly measured by probing the atomic wavefunction using the time-of-flight technique.
The effect of interactions, that unavoidably hinders localization measurements in solid-state systems, can here be reduced by going from a Bose-Einstein condensate to a dilute gas or by using Feshbach resonances.
Lastly, thanks to light-matter interaction, far-detuned laser speckles can be used as tunable disordered potentials for the atoms, allowing the exploration of the localization phase diagram.

Several experiments\cite{demarco3d,josse3d,modugno3d} on cold atomic gases have attempted to characterize metal-insulator Anderson transitions in three dimensions, 
starting from a precise measurement of the mobility edge.
This turns out to be surprisingly difficult, both theoretically and experimentally.
 Since  the Anderson transition is a second-order quantum phase transition\cite{mirlin}, the localization length diverges on the localized side 
and the diffusion coefficient slowly vanishes on the diffusive side. Distinguishing
the two behaviors requires long observation times, which is not easily achievable in experiments. 
Moreover, an atomic wavepacket expanding in a random potential inevitably contains states at different energies, thus mixing localized and extended components. For these reasons,
results on the position of the mobility edge are widely spread. 
On the theoretical side, the main difficulty comes from the fact that the metal-insulator transition takes place  in the strongly scattering regime where perturbative/diagrammatic expansions in the disorder strength fail and one has to resort to the 
indignity of numerical calculations. 
The first quasi-exact numerical estimations~\cite{ddgo} of the mobility edge turned out to lie significantly below most experimental measurements. It has been proposed that the peculiar 
anisotropic correlation functions of realistic disordered potentials, which were not taken into account in Ref.\onlinecite{ddgo}, could be at 
the origin of this discrepancy.
However, reproducing all minute details of experimental speckle configurations in numerical simulations is cumbersome, if desirable at all.
It would be more profitable to gain physical understanding of how the mobility edge changes with speckle geometry, and know its more universal features.
In the present paper, we show that the effect of anisotropy on the mobility edge can be taken into account very simply -- to a surprizingly good accuracy -- by a proper
rescaling of the characteristic energies. As most reported experimental measurements of the mobility edge do not seem to follow such scaling property, we conclude that they are actually problematic.

\section*{Statistical properties of speckle potentials} 
\begin{figure}
\includegraphics[width=1.\columnwidth]{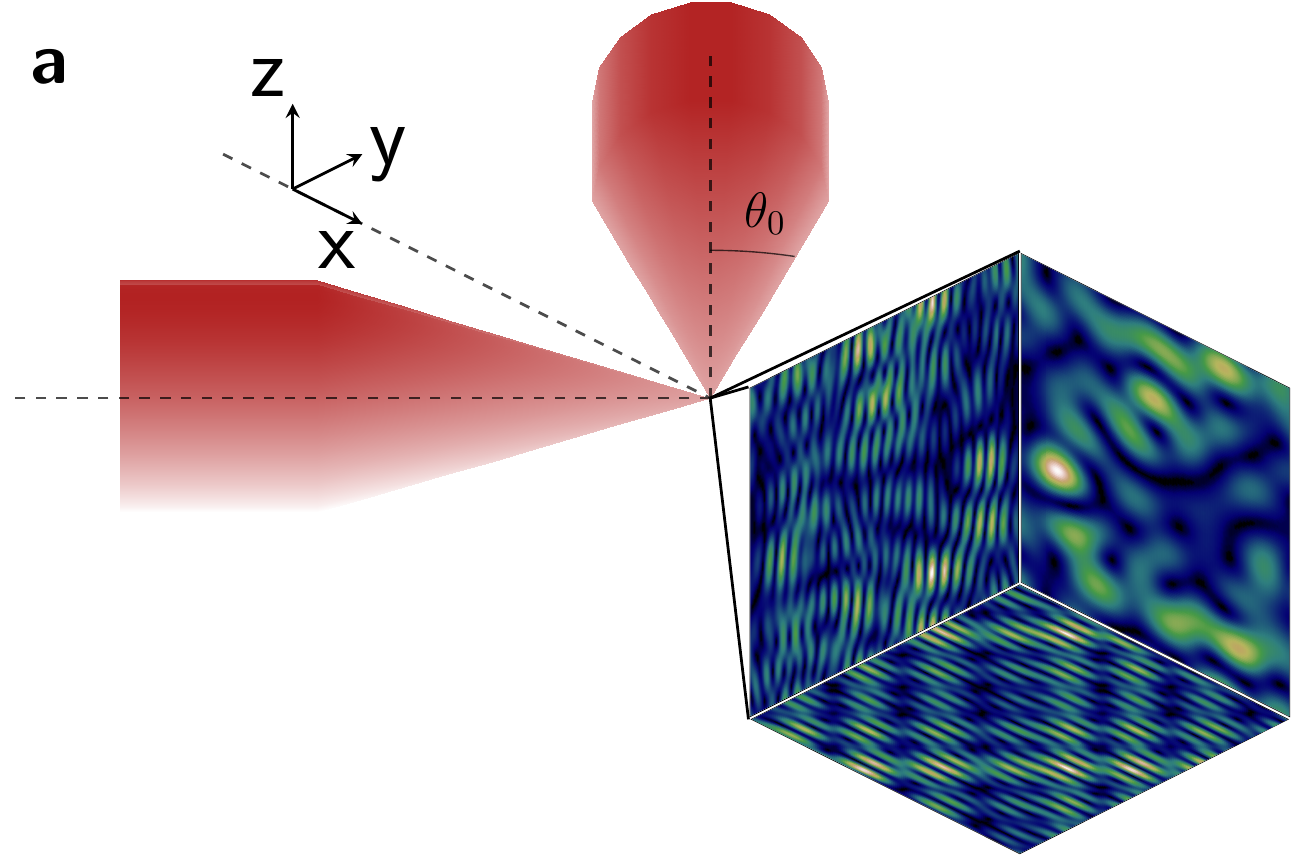}
\includegraphics[width=1.\columnwidth]{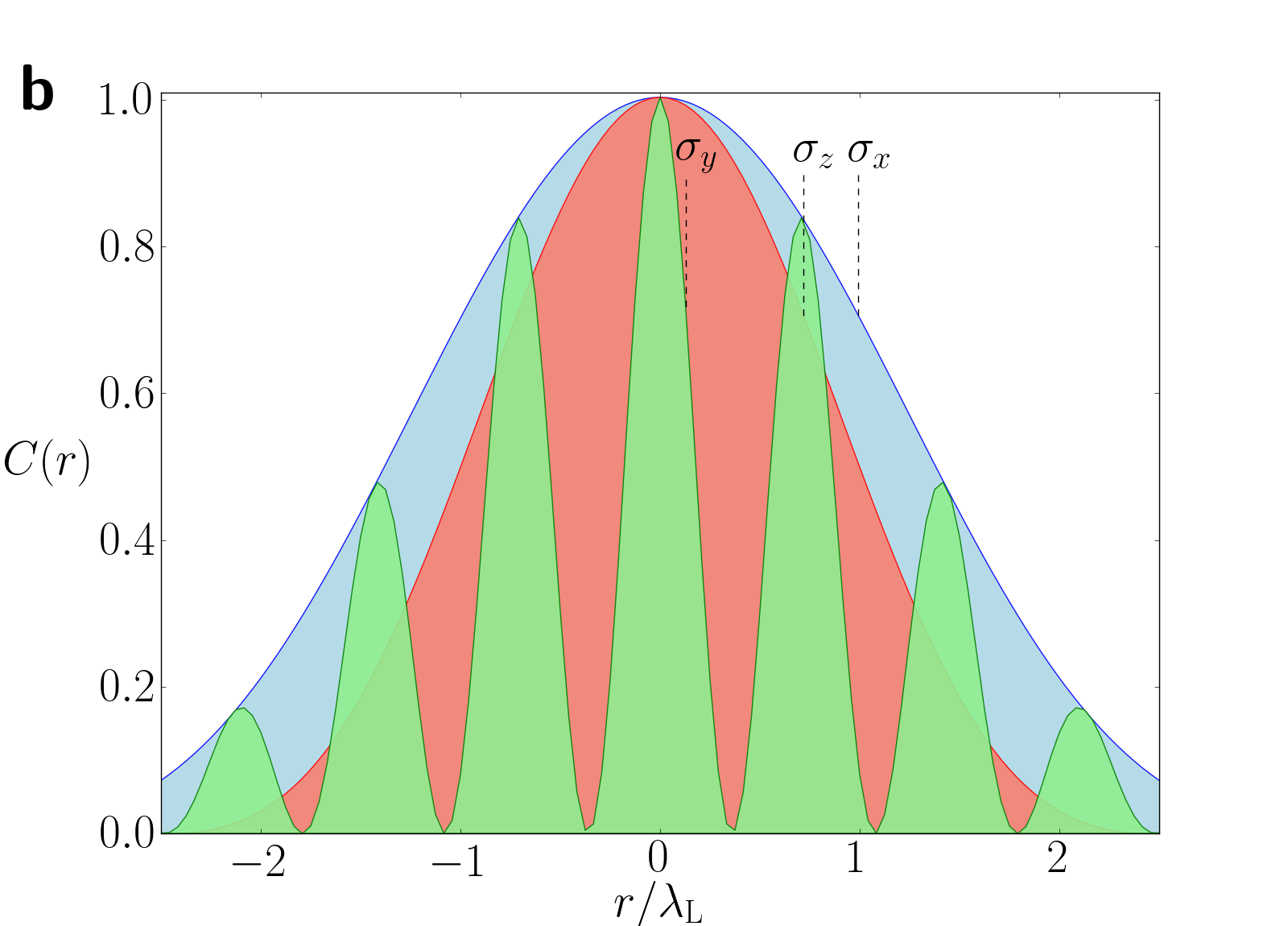}
\caption{\textbf{Properties of the optical disorder potential.} \textbf{a}, Experimental configuration for the generation of an optical disorder potential using two interfering 
coherent laser beams (along the bisectors of the $x$ and $y$ directions), as used in Refs.~\onlinecite{josse3d,modugno3d}, with same numerical aperture $\theta_0=0.3$. \textbf{b}, Two-point correlation function of the resulting disorder potential 
as function of position (in units of the laser wavelength $\lambda_L$) in the $x$, $y$ and $z$ directions. The three correlation lengths $\sigma_x$, $\sigma_y$ and $\sigma_z$ (see text) 
are markedly different in such anisotropic experimental configuration.}
\label{fig:speckle}
\end{figure}

Disordered optical potentials used in current experiments are created by shining coherent laser beams through diffusive plates,  as  sketched in Fig.~\ref{fig:speckle}.
The shift of atomic energy levels due to light-matter coupling is proportional to the intensity of radiation $I$, yielding an effective optical potential $V(\mathbf r) \propto I(\mathbf r)/\delta$ for the center-of-mass motion of the atoms, where 
$\delta$ is the detuning between the laser and atomic transition frequencies~\cite{bible}.
We restrict our study to blue-detuned speckles, corresponding to $\delta>0$, which are widely used in experiments. The optical disorder is then
always positive and its potential distribution obeys the Rayleigh law. Without loss of generality, it is customary to shift $V$ by its average value $V_0$, so that the distribution takes the form:
\begin{equation}
\label{eq:pv}
P(V) = \frac{\Theta(V+V_0)}{V_0} \exp\left(-\frac{V+V_0}{V_0}\right),
\end{equation}
$\Theta(x)$ being the unit step function. With this choice, $\langle V\rangle=0$ and the variance is $\langle V^2\rangle = V_0^2$.
The fact that the local potential distribution $P(V)$ is not Gaussian and strongly asymmetric -- in contrast
with model potentials often used in theoretical calculations --  has important consequences 
for the behaviour of the mobility edge~\cite{mpddgo}. 
Since the kinetic energy is positive, we see from Eq.~(\ref{eq:pv}) that no energy level exists below $-V_0$, implying that the mobility edge $E_c$ must satisfy the condition $E_c/V_0>-1$.

Optical speckle potentials also exhibit non-local spatial correlations. In particular, the two-point correlation function
\begin{equation}
\langle V(\mathbf{r^\prime}) V(\mathbf{r^\prime} + \mathbf{r}) \rangle = V_0^2 C (\mathbf{r}). 
\end{equation}
is characterized by finite correlation lengths which, as shown below, play a crucial role for the mobility edge.
In the simplest model of a three-dimensional isotropic speckle potential (created by a monochromatic laser of wavevector  $k_L$ 
coming from all directions of space),
the correlation function is $C(r)=[\sin(r/\sigma) / (r/\sigma)]^2$, where $\sigma = 1/k_L.$ 
For this model potential -- not easily realized experimentally -- the mobility edge has been numerically computed in Ref.~\onlinecite{ddgo}.
It was shown that the mobility edge lies \emph{way below} the average potential, $E_c<0$, a non-trivial behavior confirmed by various approximate theoretical approaches based on 
the self-consistent theory of localization\cite{Yedjour,Piraud:SCTL:PRA14}. However, all measurements reported in Ref.~\onlinecite{demarco3d}
and a subset of those appearing in Ref.~\onlinecite{modugno3d} have positive mobility edges.

In the simple case of isotropic correlations in the potential, the asymptotic behavior of the mobility edge can be deduced by considering the different energy scales of the system.
There is a natural energy scale associated with the correlation length $\sigma$, called the correlation energy\cite{Kuhn:Speckle:PRL05,Kuhn:Speckle:NJP07}:
\begin{equation}
 E_{\sigma} = \frac{\hbar^2}{m\sigma^2},
 \label{eq:correlation_energy_isotropic}
 \end{equation}
where $m$ is the atomic mass. For $V_0\ll E_{\sigma},$ the de Broglie wavelength of an atom with energy $V_0$ is much larger
than the correlation length of the potential, so that the particle can tunnel through the disorder barriers. 
This is the so-called ``quantum'' regime, where the mobility edge is expected to be very close to zero. In contrast, for $V_0\gg E_{\sigma},$ the matter wave resolves all the fine details of the disordered potential. In this ``classical'' regime, the mobility
edge is expected to be close to the percolation threshold of the potential, which is very close to $-V_0$\cite{Pilati:Bosegas:NJP10}. Among the three characteristic energy scales, $E_c, V_0, E_{\sigma},$ only their ratios matter, so that one has a unique (unknown) scaling function
such that
\begin{equation}
 \frac{E_c}{V_0}=\mathcal{F}\left(\frac{V_0}{E_\sigma}\right),
 \label{Eq:scaling_law}
\end{equation}
with $\lim\limits_{x \to 0}\mathcal{F}(x)=0^-$ and $\lim\limits_{x \to \infty}\mathcal{F}(x)\approx -1$.

The question most relevant to experiments is whether this scaling behavior applies to realistic speckles with \emph{anisotropic} correlation functions.
Two different experimental setups have essentially been developed. In Ref.~\onlinecite{demarco3d}, a single diffusive plate was used to create a speckle
pattern with a rather strong anisotropy, depending on the numerical aperture $\theta_0$ of the imaging system.
For moderate values of $\theta_0,$ the correlation lengths in the directions orthogonal and parallel to the speckle laser beam (of wavelength $\lambda_L$) are given  by $\lambda_L/\theta_0$ and $\lambda_L/{\theta_0^2}$,  respectively~\cite{goodman}.
For experimental setups in which two crossed coherent laser speckles are made to interfere~\cite{josse3d,modugno3d},
there is an additional characteristic length scale corresponding to the fringe spacing, $\lambda_L/\sqrt{2}$, generated by the two perpendicular
beams, as can be seen in Fig.~\ref{fig:speckle}.
In both cases, different correlation lengths are present in the system, so that it is no longer clear how to define the correlation energy in Eq.~\eqref{Eq:scaling_law}.

\section*{Universal scaling of the mobility edge in the quantum regime}

\begin{figure}
\includegraphics[width=1.\columnwidth]{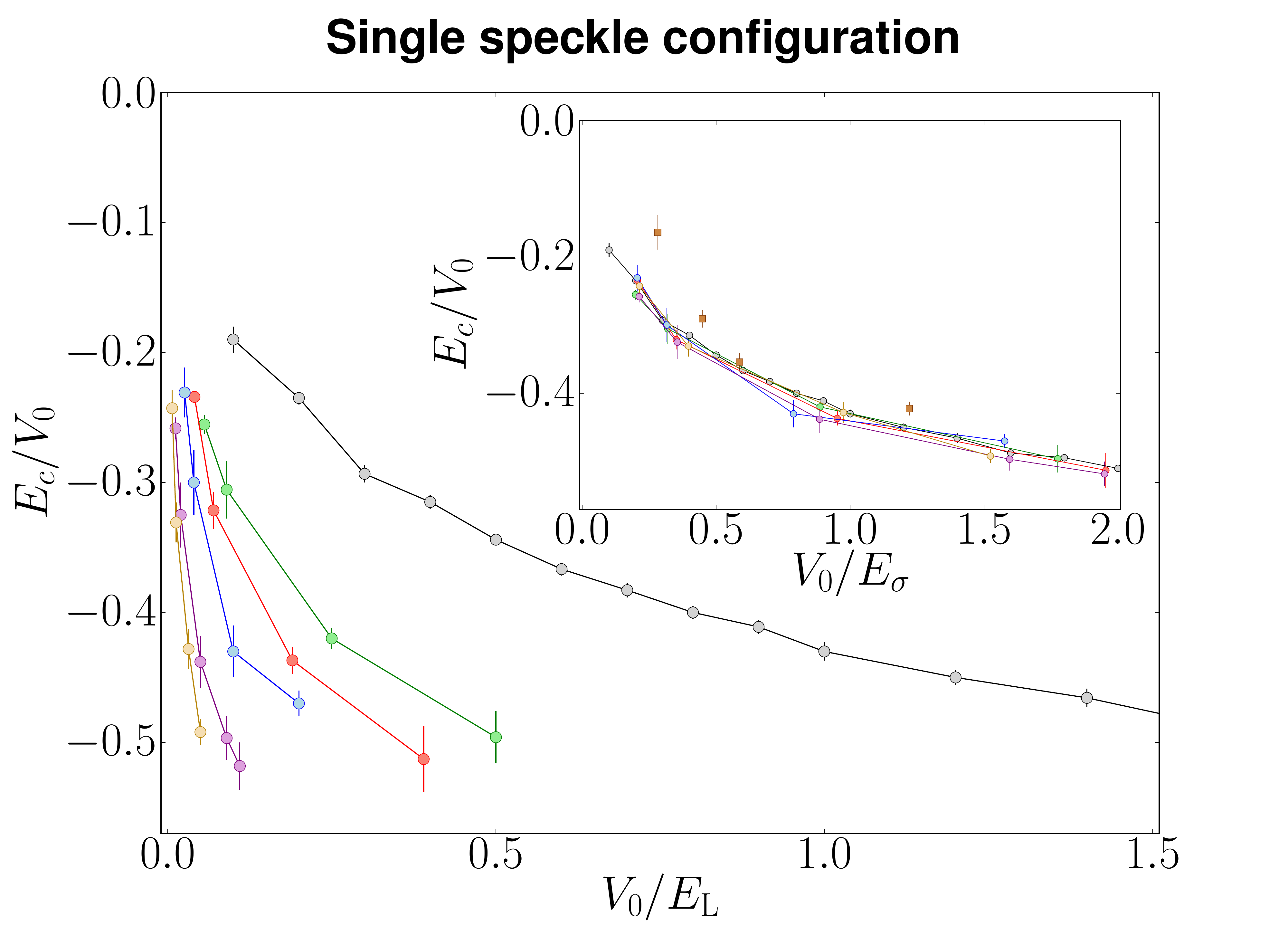}
\caption{\textbf{Mobility edge for a single laser speckle.} 
Calculated values of the normalized mobility edge $E_c/V_0$ 
of atoms in the single-speckle configuration, plotted  as a function of the disorder amplitude $V_0$ in units of $E_L=\hbar^2 k_L^2/m$
  ($k_L$ being the wavevector of the laser beam and $m$ the atomic mass). Different curves correspond to different numerical apertures $\theta_0=\{0.4, 0.5, 0.7, 0.85, 1.0\}$ (from lower to upper coloured lines) which control the degree of anisotropy. Black symbols extending to $V_0/E_L=1.5$
  correspond to the predictions of Ref.~\onlinecite{ddgo} for the isotropic case (error bars are not visible as they are smaller than symbol size). 
The inset shows that the normalized mobility edges $E_C/V_0$ versus normalized disorder amplitude 
$V_0/E_\sigma$, with average correlation energy $E_\sigma=\hbar^2/m(\sigma_\perp^2 \sigma_\parallel)^{2/3}$, 
approximately collapse to a single curve on top of the isotropic speckle result (black line) for all anisotropic configurations; $\sigma_\perp$ and $\sigma_\parallel$ being the correlation lengths in the directions orthogonal and parallel to the laser beam, respectively.  Brown squares show the results of an
independent estimation of the mobility edge\cite{Pilati:Anisotropic:PRA15}, in fair agreement with our results. }
\label{fig:sspme}
\end{figure}

\begin{figure}
\includegraphics[width=1.\columnwidth]{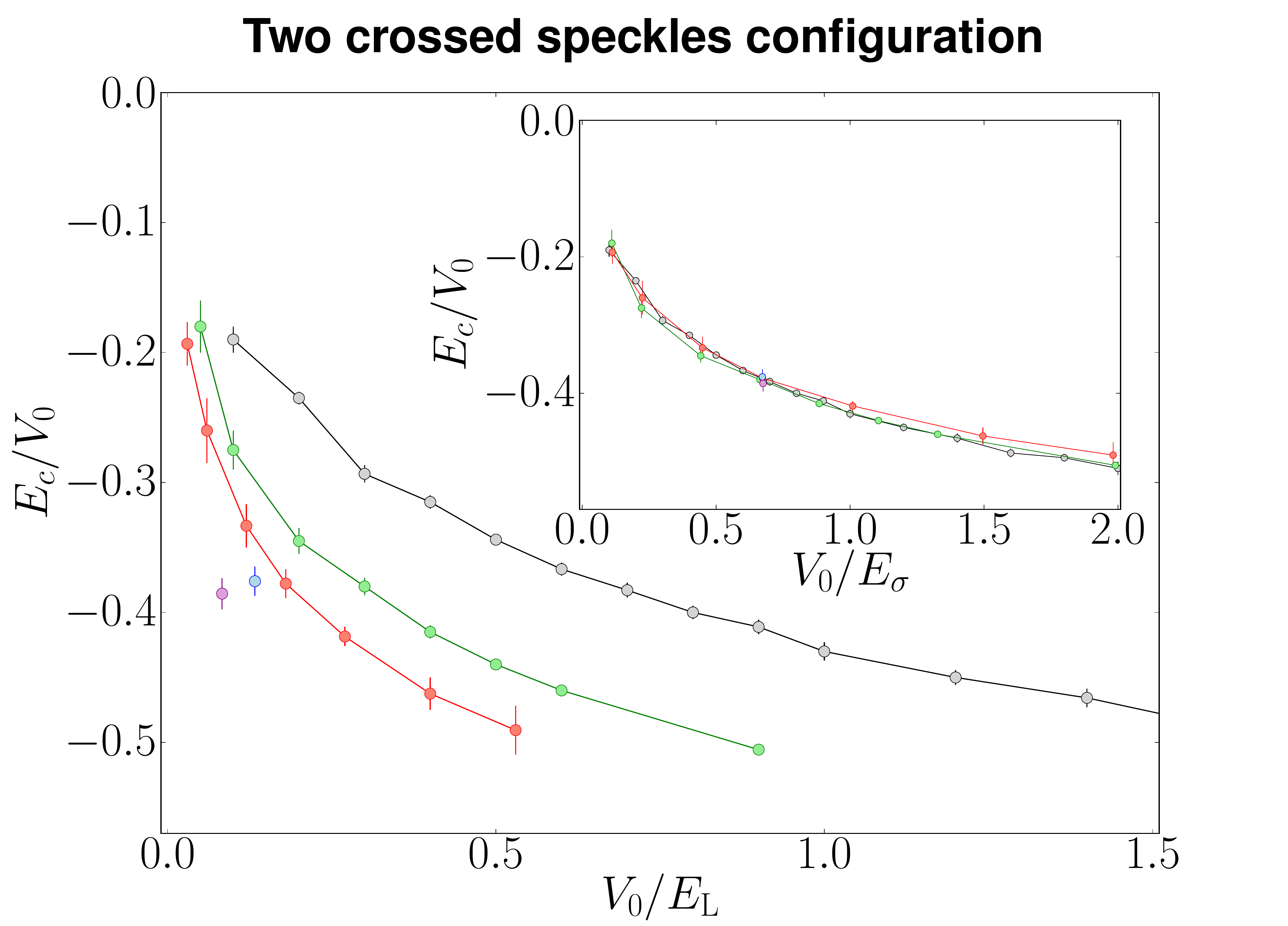}
\caption{\textbf{Mobility edge for two coherent crossed laser speckles.} Calculated values of the normalized mobility edge $E_c/V_0$ of atoms in the crossed-speckles configuration depicted in Fig.~\ref{fig:speckle}, plotted as a function of the  disorder amplitude $V_0$ in units of $E_L$ 
for two different values of the numerical aperture $\theta_0=\{0.5,0.8\}$ (respectively lower and upper coloured lines). Two additional values $\theta_0=\{0.3,0.4\}$ (respectively lower purple and upper blue point) for a given normalized disorder strength $V_0/E_\sigma=0.673$, as well as the result for isotropic speckle (black line) are also plotted.
The inset shows the same data for the normalized mobility edge $E_c/V_0$ but versus normalized disorder 
amplitude $V_0/E_\sigma$, 
with average correlation energy $E_\sigma=\hbar^2/m(\sigma_x \sigma_y \sigma_z)^{2/3}$, which collapse to a single universal curve. }
\label{fig:cspme}
\end{figure}

In order to understand how the position of the mobility edge is affected by  the anisotropy of realistic optical disorder, we have 
performed quasi-exact numerical calculations for two different speckle geometries and different values of the numerical aperture.
In Fig.~\ref{fig:sspme} we plot the normalized mobility edge $E_c/V_0$ for the \emph{single} laser speckle configuration
as a function of the disorder amplitude $V_0$, expressed in units of 
$E_L=\hbar^2 k_L^2/m$, for five different values of the numerical aperture $\theta_0$ controlling the anisotropy.
In Fig.~\ref{fig:cspme}, the same quantity is plotted for a  disordered potential generated by \emph{two} crossed identical laser speckles, for two different values of  $\theta_0$.
The first striking observation is that the mobility edge is always negative (i.e.~below the average potential $V_0$) for all anisotropic configurations,
as for the isotropic speckle whose results are also plotted in the figures.
The second striking observation is that, although the mobility edge values are widely different for different degrees of anisotropy,
the behavior in each case is strongly similar.

This suggests that the scaling law, Eq.~(\ref{Eq:scaling_law}), can be extended to the anisotropic case 
by an appropriate redefinition of the correlation energy. We thus \emph{define} the correlation energy in the anisotropic case by (see Supplementary Information):
\begin{equation}
 E_{\sigma} = \frac{\hbar^2}{m(\sigma_x \sigma_y \sigma_z)^{2/3}},
 \label{eq:correlation_energy_anisotropic}
\end{equation}
$\sigma_x, \sigma_y$ and $\sigma_z$ being the correlation lengths of the disorder along the major axes.
There is however still some ambiguity on how to define the correlation length in a given direction. Depending on the detailed
shape of the correlation function, different definitions may lead to different numerical values (up to a constant numerical
factor depending on the shape). We use a pragmatic definition, namely that $\sigma$ is the half-width at half-maximum of the central correlation peak (see Fig.~\ref{fig:speckle}b) divided by the numerical factor $\gamma\approx 1.39156$ such that $\sin(\gamma)/\gamma=1/\sqrt{2}.$ This adhoc definition ensures the compatibility with the usual definitions of the
correlation length and correlation energy in the isotropic case.

In the inset of Figs.~\ref{fig:sspme} and \ref{fig:cspme}, we replot the same numerical values of the mobility edge as function
of rescaled disorder strength $V_0/E_{\sigma}$ for the various anisotropic configurations, each configuration having a specific $E_{\sigma}(\theta_0)$ horizontal rescaling
factor as defined in Eq.~(\ref{eq:correlation_energy_anisotropic}). Amazingly, all curves collapse on a \emph{single universal} curve, independently
of the anisotropy and whether the speckle pattern is created by a single laser beam or two crossed beams.
This constitutes a major result of this paper, as it allows to take into account all complications
introduced by the inherent anisotropy of experimental speckle disorder through a simple rescaling of the correlation energy.

Strictly speaking, this universality cannot be exact, as details of the disorder correlation function must have
an influence on the position of the mobility edge. However, this influence is actually rather moderate\cite{ddgo}.
Indeed, the system being strongly scattering at the mobility edge (according to the Ioffe-Regel criterion, the mean free-path is shorter than the de Broglie wavelength), long-range correlations in the potential are essentially irrelevant. This universality is more accurate in the quantum regime $V_0\ll E_{\sigma}$ where
the matter wave averages out the disordered potential. 
In the regime of current cold-atom experiments, the universal function $\mathcal{F}$ allows to predict the position of the mobility edge within \emph{a few percent}, which is largely enough considering that the discrepancy between experimental results and theoretical predictions is of the order of 
100\% or higher.

Recently, the position of the mobility edge in the single speckle configuration has been estimated in Ref.~\onlinecite{Pilati:Anisotropic:PRA15}, using a different method based on the statistical properties
of the energy spectrum. We have added the corresponding points in the inset of Fig.~\ref{fig:sspme}.  
Although they lie slightly above our results, these data confirm that the mobility edge is very significantly below the
average potential, even in the anisotropic case.

\section*{Discussion of experimental results} 
\begin{figure}
\includegraphics[width=1.\columnwidth]{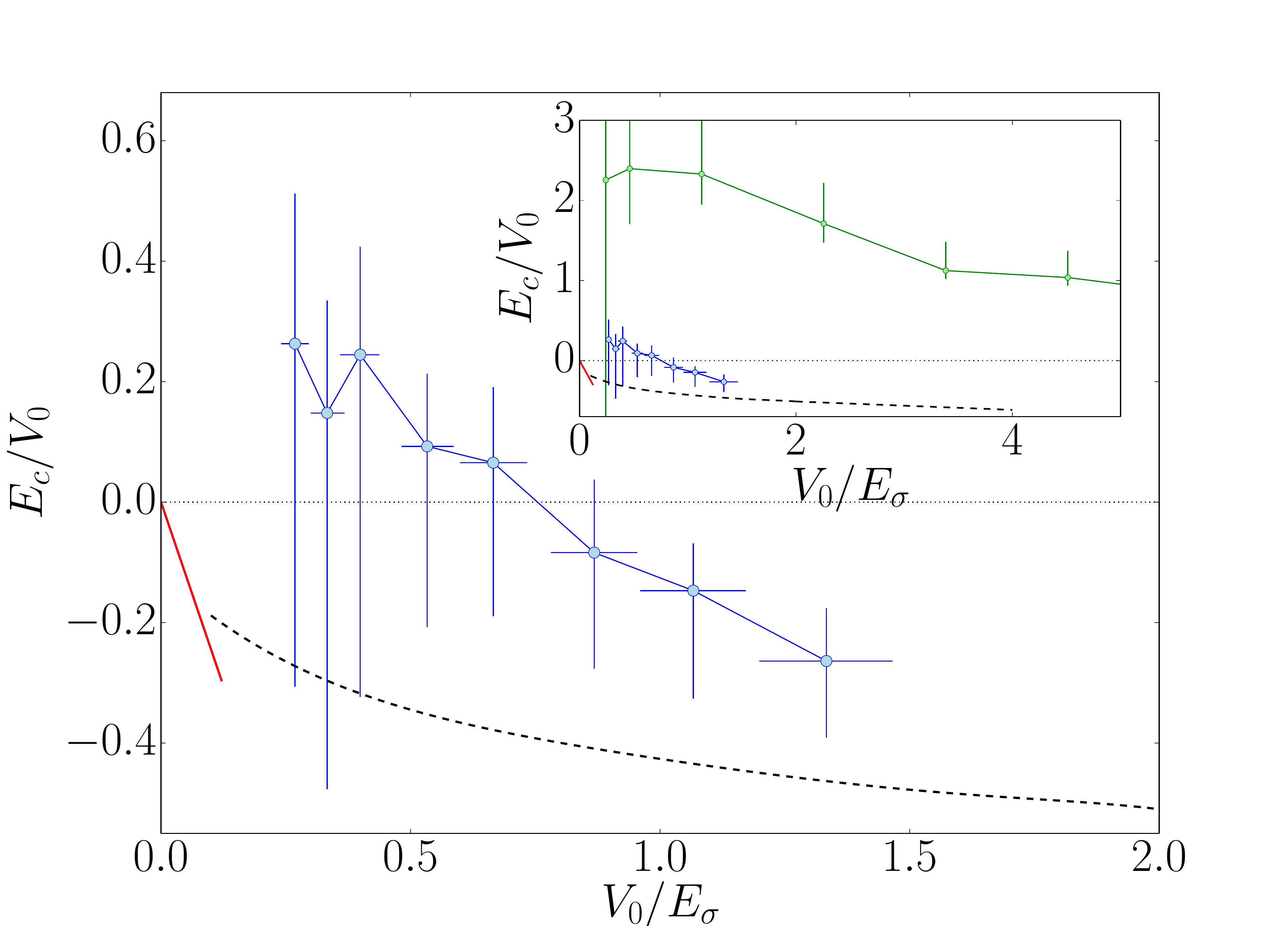}
\caption{\textbf{Experimental results.} Experimental values of the mobility edge for non-interacting cold atoms in blue-detuned speckle disorder inferred from three different experiments.
Both Ref.~\onlinecite{josse3d} (red line) and Ref.~\onlinecite{modugno3d} (blue line, with error bars) use two crossed coherent speckles, while the setup of Ref.~\onlinecite{demarco3d} uses a single, 
strongly anisotropic, speckle (green line in the inset, with error bars). 
The correlation energy $E_\sigma$ has been computed using the correlation lengths provided in each article. 
The universal curve for the mobility edge is also represented for reference (black dashed line).}
\label{fig:expme}
\end{figure}

To date, three different cold-atom experiments have reported measurements of the mobility edge in blue-detuned speckle disorder~\cite{demarco3d,josse3d,modugno3d}, 
by monitoring the temporal expansion of a localized matter wavepacket in the presence of disorder.
Although all three experiments do not use the same setup for the generation of the optical disorder potential and hence have different potential correlation functions, 
they can be quantitatively compared in the quantum regime using the above universal scaling of the mobility edge.
The experiment of Ref.~\onlinecite{demarco3d} used a very anisotropic single speckle configuration. The use of spin-polarised fermionic atoms allowed to have a relatively well defined energy of the atoms, which could be an important advantage for pinpointing the precise location of the mobility edge. 
However, the duration of the experiment may have been too short to detect slow diffusion above the mobility edge\cite{mullshap}, thus leading to a possible overestimation of the lowest diffusive energy. 
As can be seen in the inset of Fig.~\ref{fig:expme}, their inferred mobility edge is indeed much higher than other experimental and numerical estimates. Moreover, it is always above the average potential, in contradiction with all numerical and theoretical
predictions. We thus conclude that these results are impaired by some severe imperfections.

The experiment of Ref.~\onlinecite{josse3d} used two crossed speckles to achieve a more isotropic configuration, but suffered from a broad initial energy distribution of the atoms -- extending
on both sides of the mobility edge --  making a direct
measurement of the mobility edge rather difficult. An indirect estimate was given  
assuming that the mobility edge $E_c$ scales quadratically with the disorder amplitude, $E_c=\beta V_0^2/E_L.$
A ``naive'' implementation of the self-consistent theory of localization~\cite{Kuhn:Speckle:NJP07} combined
with the estimated energy distribution of the atoms gave a prediction for the localized fraction vs. disorder strength
which was not in good agreement with the experiment. However, a fair agreement was observed by correcting the prediction of the self-consistent approach by a ``heuristic shift'' quadratic in $V_0.$
This yields the straight red line $E_c/V_0=-2.44 V_0/E_{\sigma}$ shown in Fig.~\ref{fig:expme}, in decent agreement with
our numerical predictions and notably below the average potential. Unfortunately, these measurements were limited to a disorder strength much smaller than the correlation energy $E_{\sigma}$. 

The more recent experiment of Ref.~\onlinecite{modugno3d} used a similar two-speckles configuration,
with a resulting large localized fraction of atoms thanks to a better control of atom-atom interactions via  a Feshbach resonance.
When the disorder amplitude is rescaled by the appropriate $E_\sigma$, the qualitative behavior of the mobility edge with $V_0$ is similar to our numerical results, 
but quantitatively too high.
The origin of this deviation is not entirely clear to us at this moment. We note, however, that the energy distribution of atoms has not been directly measured in the experiment, but rather inferred from numerical simulations of fairly small systems. An error in this non-trivial reconstruction process could have biased their estimation of the mobility edge. 
Moreover, in these crossed-speckle configurations  -- as can be seen in Fig.~\ref{fig:speckle}.b -- small numerical apertures result in multiple correlation peaks in the direction of fringes ($y$ direction in the aforementioned figure); this could lead to an effective correlation energy smaller than the one we define in Eq.~(\ref{eq:correlation_energy_anisotropic}), hence pushing the mobility edge below the ``universal'' curve.

\section*{Conclusions}
We computed numerically the mobility edge for non-interacting cold atoms in three-dimensional speckle potentials, taking into account the long-range
anisotropic  disorder correlations that characterize realistic experimental configurations.
In the quantum regime of low disorder, which is relevant for most experiments, it was shown that the mobility edge displays a robust scaling property when the anisotropy and/or speckle geometry is varied, with the correlation energy of the potential -- defined from the geometric mean of the correlation lengths -- as the scaling parameter. 
Early experimental studies of 3D Anderson localization\cite{demarco3d,josse3d} as well as more accurate measurements\cite{modugno3d} show significant discrepancies from this universal behavior, calling for further studies. This work opens the way for more quantitative comparisons between current cold-atom experiments on 3D Anderson localization and theory/numerics.

\section*{Supplementary Information}

\textbf{Definition of the correlation energy for anisotropic speckles.}  For isotropic disorder, the characteristic energy scale separating the low-energy ``quantum'' regime 
from the high-energy ``classical'' regime is the correlation energy in Eq.~(\ref{eq:correlation_energy_isotropic}). For 
anisotropic disorder, it is customary~\cite{modugno3d,demarco3d} to define the correlation energy using
the geometrical average of the correlation lengths along the three directions, Eq.~(\ref{eq:correlation_energy_anisotropic}). Our numerical results confirm that this is indeed the energy scale relevant
for the mobility edge, and we give below two complementary arguments justifying this assumption. 

First, at very low disorder strength, the mobility edge is also at very low energy, in fact only slightly below the average potential. Then, the typical de Broglie wavelength $\lambda_\mathrm{dB}$ of the atom is larger than
the correlation lengths of the potential, so that the particle averages out all short-range fluctuations of the potential landscape. The typical size of a potential spot is the product $\sigma_x\sigma_y\sigma_z.$ The quantum wave will thus average about $N=\lambda_\mathrm{dB}^3/\sigma_x\sigma_y\sigma_z$ uncorrelated potential spots. 
From the central limit theorem, it will result in an effective renormalized potential with Gaussian fluctuations of typical strength $V_0/\sqrt{N}$ and \emph{isotropic} correlation length $\lambda_\mathrm{dB}.$  
At this stage, the details of the
correlation function of the initial potential are irrelevant, leaving only the geometric average
$(\sigma_x\sigma_y\sigma_z)^{1/3}$ as a relevant parameter. Calculations performed for specific models\cite{Piraud:Anisotropic:NJP13} confirm that the dynamics at very low energy tends to be isotropic.

A second justification comes from the calculation of the weak localization correction to classical diffusive transport. The starting point is the modification of the classical Boltzmann diffusion
picture by quantum interference due to closed loops. In the anisotropic case, where one must use a diffusion tensor,
the correction to the Boltzmann tensor $\DB$ is~\cite{Woelfle:PRB1984,Piraud:SCTL:PRA14}:
\begin{equation}
 \DD(\omega,E) = - \frac{\DB(E)}{\hbar \pi \rho(E)} \int{\frac{\rmd \vq}{(2\pi)^3}\frac{1}{-i\omega + \vq.\DB(E).\vq}} 
\end{equation}
where $\rho(E)$ is the average density of states per unit volume.
The integral has a large $\vq$ divergence which can be regularized with a convenient cut-off. The simplest choice for this momentum cut-off
is to choose along each direction $i=x,y,z$ the inverse of the transport mean free path $\ell_i$. As discussed in Refs.~\onlinecite{Piraud:Anisotropic:EPL12,Piraud:Anisotropic:NJP13}, the anisotropy of the diffusion tensor essentially reflects
the anisotropy of the transport mean free path, so that the transport mean free time $\tau$ is essentially
identical along all directions.  
The diagonal element of the diffusion tensor along direction $i$ is then $D_{ii}\propto \tau^{-1} \ell_i^2.$ 
With this choice of cut-off, the integral can be simply performed, and the correction scales like\cite{Bhatt:Anisotropic:PRB85}
$[\mathrm{det}\DB]^{-1/2}.$ In turn, this implies that, instead of the usual isotropic Ioffe-Regel criterion
$k\ell\approx 1$ for the determination of the mobility edge, we have $k[\ell_x\ell_y\ell_z]^{1/3}\approx 1.$
As, in any case, the transport mean free path in a given direction is always larger than the correlation length of the disorder
in the same direction, we end up again with the conclusion that the relevant length scale is $(\sigma_x\sigma_y\sigma_z)^{1/3}.$ 

Note however that the weak localization correction is here predicted to
have no effect on the anisotropy of the diffusion tensor\cite{Piraud:Anisotropic:EPL12}, but only leads to a global
renormalization. More sophisticated approaches~\cite{Yedjour,Piraud:SCTL:PRA14} as well as numerical
results~\cite{Zambetaki:PRB1997} show that it is not the case. Close to the mobility edge,
transport is less anisotropic than in the classical Boltzmann regime. A quantitative understanding of this phenomenon
is highly desirable, but beyond the scope of this paper. Preliminary results indicate that some anisotropy remains
at the mobility edge. In any case, it does not seem to invalidate that the anisotropic correlation energy Eq.~(\ref{eq:correlation_energy_anisotropic}) is the relevant energy scale of the problem.

\section*{Methods}

\textbf{Numerical generation of the disordered potential.}  In order to reproduce the correct disorder correlation functions
-- i.e. the ones used in real experiments -- we generate numerically a ``random'' electromagnetic field by combining
many plane waves with different wavevectors and uncorrelated random phases. Because a monochromatic laser is used to create the speckle,
the wavevectors (of unit modulus) are distributed on the surface of a sphere, typically with a Gaussian distribution like in most experiments, 
inside a single cap with half-angle $\theta_0$ (in the ``single speckle'' configuration) or two perpendicular caps (in the ``crossed speckles'' configuration).
All wavevectors are taken with the same polarization. The disorder at a given position is then simply the modulus square
of the electric field computed by coherently combining the various plane waves.
We carefully checked that we obtain in the small $\theta_0$ limit the correlation functions predicted in Refs.~\onlinecite{goodman,Piraud:Anisotropic:NJP13}. 

\noindent \textbf{Transfer-matrix method.} The mobility edge is calculated using the same procedure outlined in Ref.~\onlinecite{ddgo}. Starting from a spatial discretization of the 3D Schr\"odinger equation on a grid, whose spacing $\Delta$  is
much smaller than the correlation lengths of the potential and de Broglie wavelength (we used $\Delta\sim\lambda_L/6$ for all computations), we end up with an effective Anderson model 
with hopping term $J=\hbar^2/(2m \Delta^2)$ and correlated on-site disorder.
We then proceed in the standard way \cite{kinnonkramer1983}
by computing  recursively the total transmission of a bar-shaped sample of width $M$ and length $L$,
such that $L \gg M,$ at fixed energy $E$ and disorder amplitude $V_0$. 
As quasi-1D systems  are always exponentially localized, the transmission decays exponentially with $L$
 over 
the localization length $\lambda_M(E,V_0).$
Using the finite-size scaling technique on the adimensional localization length $\lambda_M(E,V_0)/M$ allows one to probe the 3D Anderson transition, 
see Ref.~\onlinecite{kinnonkramer1983} for details.
Compared to the isotropic case computed in Ref.~\onlinecite{ddgo}, the difficulty lies in the anisotropic correlation function, requiring significantly 
larger system sizes -- especially along the directions of largest correlation lengths -- to reach convergence. Typically, we used transverse sizes of up to $M=96$ sites for the finite-size scaling analysis and lengths around $10^6$ sites. Small values of $\theta_0,$ as used in Ref. \onlinecite{modugno3d}, are thus very challenging numerically; more than two million hours of CPU time on supercomputer resources have been used to obtain the results of Figs.~\ref{fig:sspme} and \ref{fig:cspme}.

\vspace{1cm}

\section*{Acknowledgements}
\begin{small} 
We acknowledge discussions with V. Josse and thank S.~Pilati for sending us the raw data of Ref.~\onlinecite{Pilati:Anisotropic:PRA15}. M. Pasek  was supported by ERC (Advanced Grant ``Quantatop'')  and the Region Ile-de-France  in the framework of DIM Nano-K (project QUGASP). 
The authors were granted access to the HPC resources of TGCC under the allocations 2014-057301, 2015-057301, 2015-057083, 2016-057629 and 2016-057644 made by GENCI (``Grand Equipement National de Calcul Intensif'').
\end{small}

\end{document}